\pdfoutput=1
\documentclass[submission]{IEEEtran}
\usepackage[utf8]{inputenc}

\usepackage{siunitx}
\usepackage{multirow}
\usepackage{todonotes}
\usepackage{setspace}
\usepackage{graphicx}
\usepackage{pifont}
\newcommand{\cmark}{\ding{51}}%
\newcommand{\xmark}{\ding{55}}%
\usepackage{subcaption}

\usepackage{adjustbox}
\usepackage{multirow}
\usepackage{amsmath,amssymb,amsfonts, amsthm}
\usepackage[linesnumbered,ruled]{algorithm2e}
\usepackage{tabularx}
\newcolumntype{Y}{>{\centering\arraybackslash}X}
\usepackage{algpseudocode}
\usepackage[english]{babel}

\usepackage{tabularx}
\usepackage[nameinlink,capitalise]{cleveref}

\begin{document}

\title{Automatic Hardware Trojan Insertion using Machine Learning}
\author{\IEEEauthorblockN{Jonathan Cruz$^{1}$, Pravin Gaikwad$^{1}$,
Abhishek Nair$^{2}$, 
Prabuddha Chakraborty$^{1}$,
and Swarup Bhunia$^{1}$\\
}
\IEEEauthorblockA{$^{1}$Department of Electrical and Computer Engineering, University of Florida, Gainesville, FL, USA\\
$^{2}$Department of Electrical Engineering, Stanford University, Stanford, CA, USA\\
Email: \{jonc205,  pravin.gaikwad,p.chakraborty\}@ufl.edu,\\ aanair@stanford.edu, swarup@ece.ufl.edu}}

\maketitle

\begin{abstract}

Due to the current horizontal business model that promotes increasing reliance on untrusted third-party Intellectual Properties (IPs), CAD tools, and design facilities, hardware Trojan attacks have become a serious threat to the semiconductor industry. Development of effective countermeasures against hardware Trojan attacks requires: (1) fast and reliable exploration of the viable Trojan attack space for a given design and (2) a suite of high-quality Trojan-inserted benchmarks that meet specific standards. The latter has become essential for the development and evaluation of design/verification solutions to achieve quantifiable assurance against Trojan attacks. While existing static benchmarks provide a baseline for comparing different countermeasures, they only enumerate a limited number of handcrafted Trojans from the complete Trojan design space. To accomplish these dual objectives, in this paper, we present MIMIC, a novel AI-guided framework for automatic Trojan insertion, which can create a large population of valid Trojans for a given design by mimicking the properties of a small set of known Trojans. While there exist tools to automatically insert Trojan instances using fixed Trojan templates, they cannot analyze known Trojan attacks for creating new instances that accurately capture the threat model. MIMIC works in two major steps: (1) it analyzes structural and functional features of existing Trojan populations in a multi-dimensional space to train machine learning models and generate a large number of ``virtual Trojans" of the given design, (2) next, it binds them into the design by matching their functional/structural properties with suitable nets of the internal logic structure. We have developed a complete tool flow for MIMIC, extensively evaluated the framework by exploring several use-cases, and quantified its effectiveness to demonstrate highly promising results.

\end{abstract}
\begin{IEEEkeywords}Machine Learning, Hardware Trojans, Automated Trojan Insertion
\end{IEEEkeywords}

\section{Introduction}
Trustworthiness of hardware has emerged as a primary concern for modern electronics. The rapid adoption of the horizontal business model by semiconductor companies has led to increased incorporation of third-party intellectual property (3PIP) blocks, electronic design automation tools and commercial off-the-shelf (COTS) components into electronic hardware, which are often untrusted. Security verification of hardware has already emerged as a challenging problem as hardware designs have become more complex.  The presence of untrusted entities in the hardware life cycle, which introduces the threat of malicious hardware alterations or Trojan attacks, further exacerbates this problem \cite{ref:mero} \cite{hoque2018hardware}.  An adversary can use hardware Trojan attacks for various malicious purposes, which include leakage of critical information (e.g., crypto key), access control violation, and Denial of Service (DoS) during field operation.  Security verification engineers will face extreme difficulty trying to observe and detect hardware Trojans as they are stealthy by construction and made to bypass traditional design verification.  

The emerging threat of hardware Trojan attacks has motivated researchers to create effective Trojan detection and prevention methods. However, the development and evaluation of countermeasures against hardware Trojans require a suite of Trojan-inserted benchmarks. Additionally, to quantify the assurance provided by a countermeasure for a given design, a designer requires the capability to rapidly explore the Trojan attack space for a given design.  Previously, researchers had crafted custom Trojans -- however custom Trojans do not lend themselves to comparing against other techniques.  A suite of 96 Trojan inserted benchmarks across 16 unique designs currently exist on Trust-hub, but are static~\cite{shakya2017benchmarking}  -- a fixed and small subset of the evolving hardware threat.  The Trojan space is sufficiently large and evolving and cannot be accurately captured with a limited number of benchmarks.  Cruz et al. introduced a methodology for automated Trojan benchmarking, which addresses several of the aforementioned problems \cite{ref:trit}.  Nevertheless, previous automated benchmarking tools rely on only one feature for Trojan insertion -- signal probability. While this will guarantee low triggerability through logic testing, such a process limits the potential Trojan space by only looking through a one-dimensional lens. We cannot expect to capture the behavior of a Trojan in a design under such assumptions. Moreover, as designs are getting more complex, the space in which an adversary can insert a stealthy Trojan similarly becomes large and more complex. Current Trojan insertion tools do not adequately utilize the growing design complexity or reliably explore the complex Trojan attack space for a given design. 

 \begin{figure*}[t!]
 	\centering
 	\includegraphics[width=\textwidth]{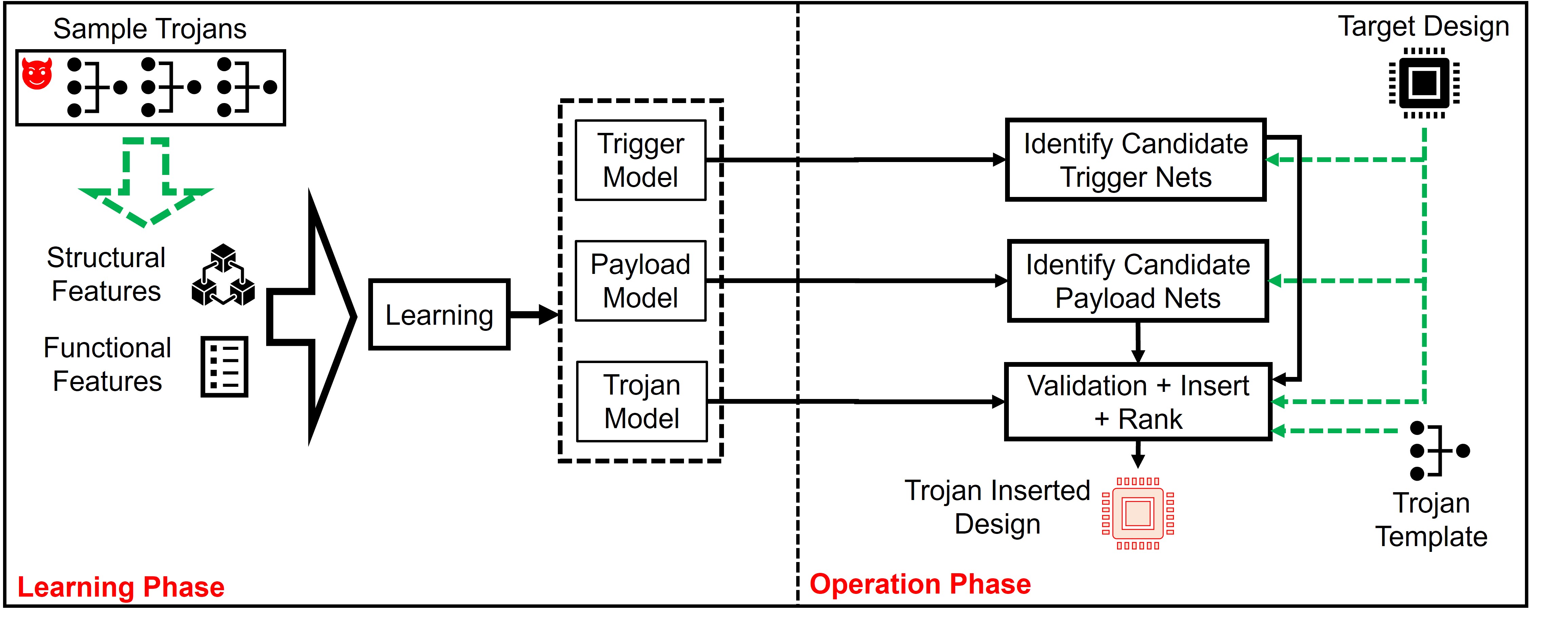}
 	\caption{The major steps and overall flow of the MIMIC framework.}
 	\label{MIMIC_flow}
 \vspace{-0.05in}
 \end{figure*}

Motivated by these opportunities, we propose \textbf{MIMIC} (\underline{M}achine \underline{I}ntelligence based \underline{M}alicious \underline{I}mplant \underline{C}reation) a machine learning-based framework for automatically generating Trojan benchmarks. \cref{MIMIC_flow} illustrates the overall flow of the proposed framework.  The MIMIC framework is split into two main phases: 1) the learning phase and 2) the operation phase.  For learning, MIMIC extracts features from a pool of existing Trojans and uses these features to train machine learning (ML) models for identifying trigger nets, payload nets, and capturing Trojan behavior.  The operation phase is where MIMIC inserts Trojans. MIMIC expects a flattened Verilog netlist. The tool then identifies candidate nets using the Trigger and Payload models.   While the algorithm selects a set of nets based on their fitness determined by the ML models, the set of nets may never simultaneously achieve all rare values or may form an unintended combinational loop, creating an invalid Trojan. Therefore, each set of nets is verified both functionally and structurally to ensure only valid Trojans are inserted. We introduce the notion of ``virtual Trojans'' or the Trojan feature set (behavior) extracted from each set of realized candidate trigger and payload nets.  These virtual Trojans are then compared to a reference Trojan feature vector sampled from the Trojan behavior ML model.  The virtual Trojans are sorted based on this comparison. Finally, the fittest Trojans are bound to the design using a Trojan template and suitable internal nets. Several configurations are exposed to the user to control the type of Trojan, feature weights, and models used during insertion.

Unlike previous approaches, MIMIC learns from an existing Trojan population and automatically extracts a multi-dimensional feature space consisting of structural and functional features from nets in a gate-level netlist.  MIMIC captures the complexities and provides an attacker with suitable locations for inserting a given Trojan template.
In this way, the framework tries to mimic an intelligent attacker and consider several qualities that contribute to the low detectability of stealthy Trojan.  The framework builds several machine learning models from which new Trojans of similar quality to the original population can be automatically generated and inserted.  A key advantage of our proposed framework over current tools is that all the Trojans present in the provided database informs the MIMIC Trojan insertion process. Therefore, Trojans generated by MIMIC can automatically adapt in response to new Trojan discoveries in the wild.  The flexibility of MIMIC enables users to control the number and type of Trojan inserted. We note different Trojan types (i.e., combinational, and sequential) often have diverse or even conflicting features. Therefore, the framework allows users to control feature weights and the machine learning models used for Trojan insertion, allowing for more representative modeling. 
 
In summary, we make the following contributions in the paper:
\begin{itemize}
    \item We identify the existing problems of current Trojan benchmarking tools, which do not adequately encapsulate the Trojan space for a given design.
    \item We introduce MIMIC, a machine learning-based flexible hardware Trojan insertion framework that can create a large population of feasible and high-quality Trojan attacks for a given design by mimicking the properties from a small set of known Trojans. MIMIC can extract relevant structural and functional features from a set of attack instances to capture Trojan behavior. MIMIC enables us to reliably explore the Trojan attack space for a given design with user-configurable attack features (e.g., Trojans of specific types of trigger or payload). To the best of our knowledge, MIMIC is the first automated Trojan insertion framework that uses both supervised and generative machine learning.
    \item We develop a complete framework, including associated algorithms and tools (e.g., automatic feature extraction, Trojan insertion into a design, and validation of an inserted Trojan) and their integration with widely-used ML algorithms. We provide the necessary details on this framework in the paper. 
    \item Based on the algorithms presented, our implementation can insert diverse Trojan types in a target design by leveraging the Trojan behavior captured using a set of machine learning models.
    \item We evaluate MIMIC on four ISCAS89 benchmarks using four different Trojan templates and quantify its effectiveness. We extensively analyze our framework's ability to mimic a given set of Trojans in various scenarios. 
    \item Finally, we study the efficacy of transfer learning by varying the Trojan templates used during testing.
\end{itemize}

 The rest of the paper is organized as the following: \cref{background} describes the preliminaries and previous efforts towards automated Trojan benchmarking. \cref{motivation} discusses why automated benchmarking of Trojans is needed and observations with current tools that motivate MIMIC. \cref{methodology} describes the machine learning-based Trojan insertion process. Section \ref{results} presents an analysis of MIMIC's ability to reproduce similar Trojans under several circumstances. Further discussion and future research directions with a conclusive summary are presented in Section \ref{conclusion}.

\section{Background and Related Works}
\label{background}
The proposed methodology can create extensive benchmarks using machine intelligence for developing countermeasures for hardware Trojan attacks.  
In this section, we provide a brief introduction to hardware Trojans. We also present existing works on automated Trojan benchmarking tools.
\subsection{Hardware Trojans}
Hardware Trojans are deliberate design modifications that create unwanted functionality outside of the defined specifications.  These malicious functions can range from leaking information and malfunctioning during critical operations to bypassing security measures \cite{ref:taal}.  Hardware Trojans can be placed in several abstractions of a design throughout its lifecycle: soft IP (i.e., RTL), firm IP, gate-level netlist, or in the IC by an untrusted foundry.  At the most basic level, Trojans consist of a trigger and payload sub-circuit. The trigger sub-circuit is used to activate the Trojan under specific conditions so as to avoid detection. Trigger circuits generally take input signals from the design and ensure an unlikely activation sequence reducing the chance of accidental activation or discovery. The payload sub-circuit is responsible for carrying out the malicious function through functional output, side-channel, or parametric values(i.e.: temperature, current). In the case of always-on Trojans, the trigger circuit does not exist, but the payload must carry out non-functional effects to remain stealthy \cite{lin2009moles}.

\subsection{Automated Trojan Insertion}

\begin{table*}[t!]
\centering
\caption{Comparison of Proposed Framework with Existing Trojan Benchmarking Tools}
\label{table:related}
\begin{tabular}{|c|c|c|c|c|}  \hline
  \multicolumn{1}{|c|}{\textbf{Tool}} & \multicolumn{1}{|c|}{\textbf{Domain}} & \multicolumn{1}{|c|}{\textbf{Automation}}  & \multicolumn{1}{|c|}{\textbf{Features for Insertion}} & \multicolumn{1}{|c|}{\textbf{Learning}}           \\ \hline
  HAL~\cite{ref:HAL} & ASIC/FPGA & \xmark & Neighborhood Control Value & \xmark \\ \hline
  TAINT~\cite{ref:taint} & FPGA & \xmark & N/A & \xmark \\ \hline
  TRIT~\cite{ref:trit} & ASIC & \cmark & Signal Probability  & \xmark  \\ \hline
  TRIT-PCB~\cite{ref:trit-pcb} & PCB & \cmark & Signal Probability & \xmark \\ \hline
  S. Yu et al.~\cite{ref:improved} & ASIC & \cmark & Transition Probability & \xmark \\ \hline
  MIMIC (Proposed) & ASIC & \cmark & Structural \& Functional Features & \cmark \\ \hline
  
\end{tabular}
\\\cmark = Included; \xmark = Excluded.
\end{table*}
 Efforts have been made towards the automatic generation of hardware Trojan benchmarks for verifying Trojan detection techniques.  TAINT is a tool for automated hardware Trojan insertion in FPGA designs \cite{ref:taint}. Users are provided the option to choose activation or trigger conditions and can select trigger and payload templates from a database of predefined Trojans.  The tool can also automatically suggest physical locations for the Trojan within the FPGA. HAL \cite{ref:HAL} is a gate-level reverse engineering tool that aids in identifying areas of interest in a digital design to insert a Trojan. The authors discussed inserting Trojans to bypass power-up self-tests and leak keys in cryptographic circuits through the use of HAL.  Both of these tools require manual effort for the insertion of Trojans \cite{ref:HAL, ref:taint} which precludes the creation of a large number of benchmarks. In \cite{ref:taint}, the user is expected to select the nets to be used as triggers based on suggestions made by the tool, while \cite{ref:HAL} is primarily a reverse engineering tool that facilitates manual inspection and selection of appropriate trigger and payload nets. Manual selection of triggers can be a daunting task due to the required step of verifying all nets that can achieve their rare values simultaneously.

TRIT \cite{ref:trit} is a proposed tool that automatically selects suitable trigger nets in a given design for inserting various templates of Trojans from a database. The trigger nets are selected by analyzing the activation probability of each net in the original design. While rare activation probability is a reasonable feature to find potential trigger nets for placing stealthy combinational and sequential Trojans, a number of other functional and structural information of the nets are ignored during the search process. Constraining to just one feature impedes the ability of the users to accurately mimic the desired Trojan model. Furthermore, insertion of other classes of Trojans (e.g., De-trust \cite{zhang2014detrust} and always-on \cite{lin2009moles}) may require different features other than activation probability. The authors in \cite{ref:improved} propose a similar flow to TRIT, but utilize transition probability instead of simulation-based signal probability.  TRIT-PCB~\cite{ref:trit-pcb} is another work inspired by TRIT, but applied at the printed circuit board abstraction.

\section{Motivation}
\label{motivation}
In this section, we describe the need for an automated Trojan attack space exploration and benchmarking tool. We also make several observations on major limitations of existing benchmarking tools that motivate the need for the MIMIC framework.
\subsection{Need for Automated Trojan Insertion}
Before the introduction of Trojan benchmarks, researchers evaluated Trojan detection schemes with handcrafted Trojan examples. Trust-hub.org web portal then introduced, about a decade ago, a suite of 96 total static Trojan benchmarks with which researchers could compare results. However, the benchmarks are limited in number, which prevents their usage in many application scenarios, such as: (1) evaluation of a countermeasure for a specific design or new and emerging Trojan attacks, which are not realized in existing Trojan benchmarks; (2) exploration of hard-to-detect Trojans of a specific type or behavior in a given design; and (3) machine learning-based trust verification framework that requires a large dataset for training.  The first two scenarios reflect use cases where a chip designer explores Trojan attack vulnerabilities for a design and verifies its robustness against Trojan attacks with known protection methods. An automated insertion tool can address these shortcomings by allowing for the creation of a large number of Trojans across any number of benchmarks. Next, we highlight several significant use-cases of MIMIC:
\begin{enumerate}
    \item \textbf{A Sampling-Based Trust Metric:} A large population of Trojan benchmarks can help us to reliably estimate Trojan coverage for a Trojan detection or trust verification method \cite{ref:mero} \cite{ref:mers}. Given that an inordinately large number of possible Trojan attacks are possible, a sampling-based trust metric can be developed that can effectively quantify the Trojan coverage for a target protection method.  
    \item \textbf{Machine Learning (ML) based Hardware Trust Verification:} The proposed framework can be used to create a large unbiased population of synthetic Trojans of specific types and inserted in specific class of designs. It enables us to create powerful ML-based hardware IP trust verification solutions, which can leverage the synthetic Trojan datasets for training and evaluation purposes. 
    \item \textbf{Explore Possible Trojan Attacks in a Design:} MIMIC enables automatic exploration of Trojan attack space for a given design. Specifically, the configurable options supported by MIMIC enable us to explore Trojan attacks of specific trigger/payload behavior for the entire or part of a design. 
    \item \textbf{Design-for-Trust Solution:} MIMIC can be used to develop a closed-loop integrated design-for-trust solution, where design solutions for protecting against Trojan attacks can be guided by the possible Trojan attacks explored by MIMIC. For example, if a design solution fails to protect a subset of synthetic Trojans, the integrated solution can iteratively incorporate additional countermeasures to protect against uncovered Trojans. 
    \item \textbf{Big-data Analytics on Test Data:} The ability of MIMIC to generate a large volume of Trojan data for various Trojan types, including emergent ones, and for a diverse set of designs is expected to fuel the application of big data analytics to enhance our understanding of Trojan vulnerabilities and the effectiveness of a protection method. 
    
\end{enumerate}
\subsection{Underestimating Trojan Behavior}
\begin{figure*}
\centering
\begin{subfigure}[b]{0.4\textwidth}
         \centering
         \includegraphics[width=\textwidth]{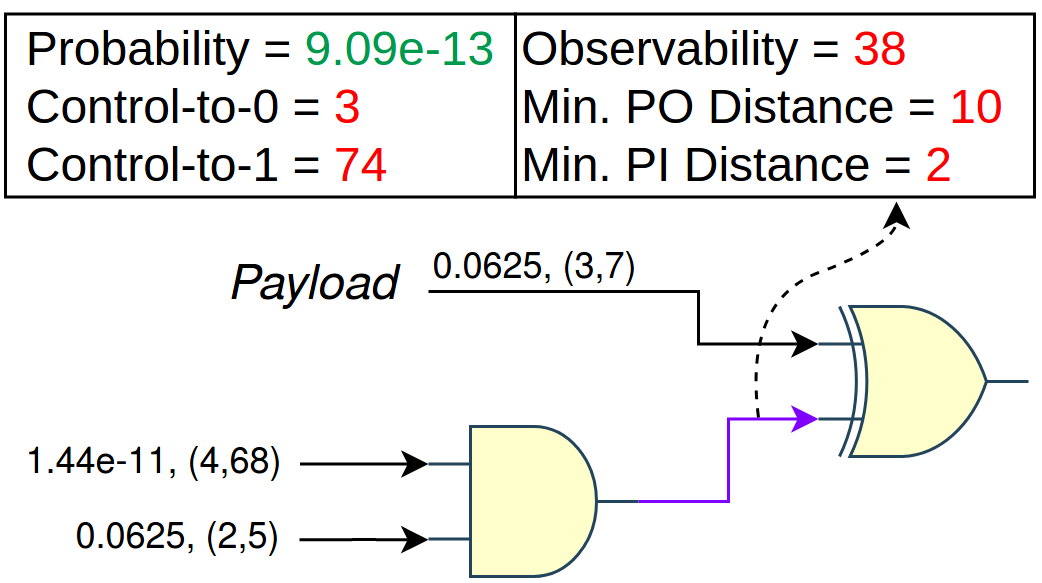}
         \caption{}
         \label{fig:motivation1}
     \end{subfigure}
     \begin{subfigure}[b]{0.4\textwidth}
         \centering
         \includegraphics[width=\textwidth]{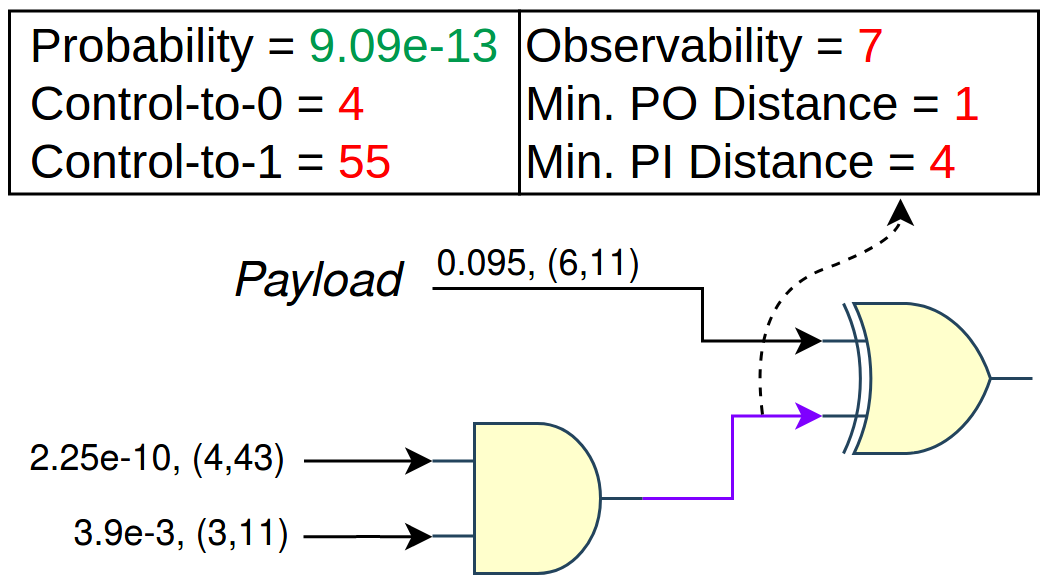}
         \caption{}
         \label{fig:motivation2}
     \end{subfigure}
    
        \caption{Two Trojans templates inserted in different locations in a design with the same activation probability. The features shown on the trigger input and payload nets are signal probability, control-to-0, control-to-1.}
        \label{fig:motivation}
\end{figure*}

Existing automated benchmarking tools rely on signal probability for inserting Trojans.  The resulting Trojans are guaranteed to be below some activation probability, but remain unconstrained with respect to other relevant features. For example, \cref{fig:motivation} shows two 2-trigger AND Trojans that can be inserted with an automated benchmarking tool. Despite exhibiting the same activation probability, the Trojans are actually very distinct when viewing them through the lens of other relevant functional and structural features.  In an ideal scenario, an attacker would look to control all relevant features of a Trojan to minimize the chance of detection, especially in light of all these ML Trojan detection schemes.  Given these observations, we can conclude that Trojans are insufficiently modeled by considering only one feature.  Therefore, the natural question becomes: How can we properly describe Trojans as a function of a multi-dimensional feature space?  Furthermore, can we learn how a Trojan behaves in a benchmark and identify similar locations? To these ends, we have developed MIMIC -- a machine learning Trojan benchmarking framework to capture and reproduce Trojan behavior.
\section{Machine Intelligence based Insertion}
\label{methodology}
In this section, we describe the methodology and framework of MIMIC. It is capable of learning the commonly occurring patterns in a class of Trojans and generating a new population of similar Trojans. The overall flow is depicted in \cref{MIMIC_flow}. 

\subsection{Learning Phase}
The framework first learns the commonly occurring patterns in a set of sample Trojans (the class we are trying to mimic). This learning is achieved through training and the creation of three ML models: a trigger model, a payload model, and a Trojan model. The trigger model and the payload model learn to identify nets in a design ideal for being trigger and payload nets, respectively. Based on the outputs of the trigger model and the payload model, the Trojan model is used to create, rank, and insert Trojans in the target design.

\subsubsection{Feature Extraction}

\begin{figure}[b]
	\centering
	\includegraphics[width=\columnwidth]{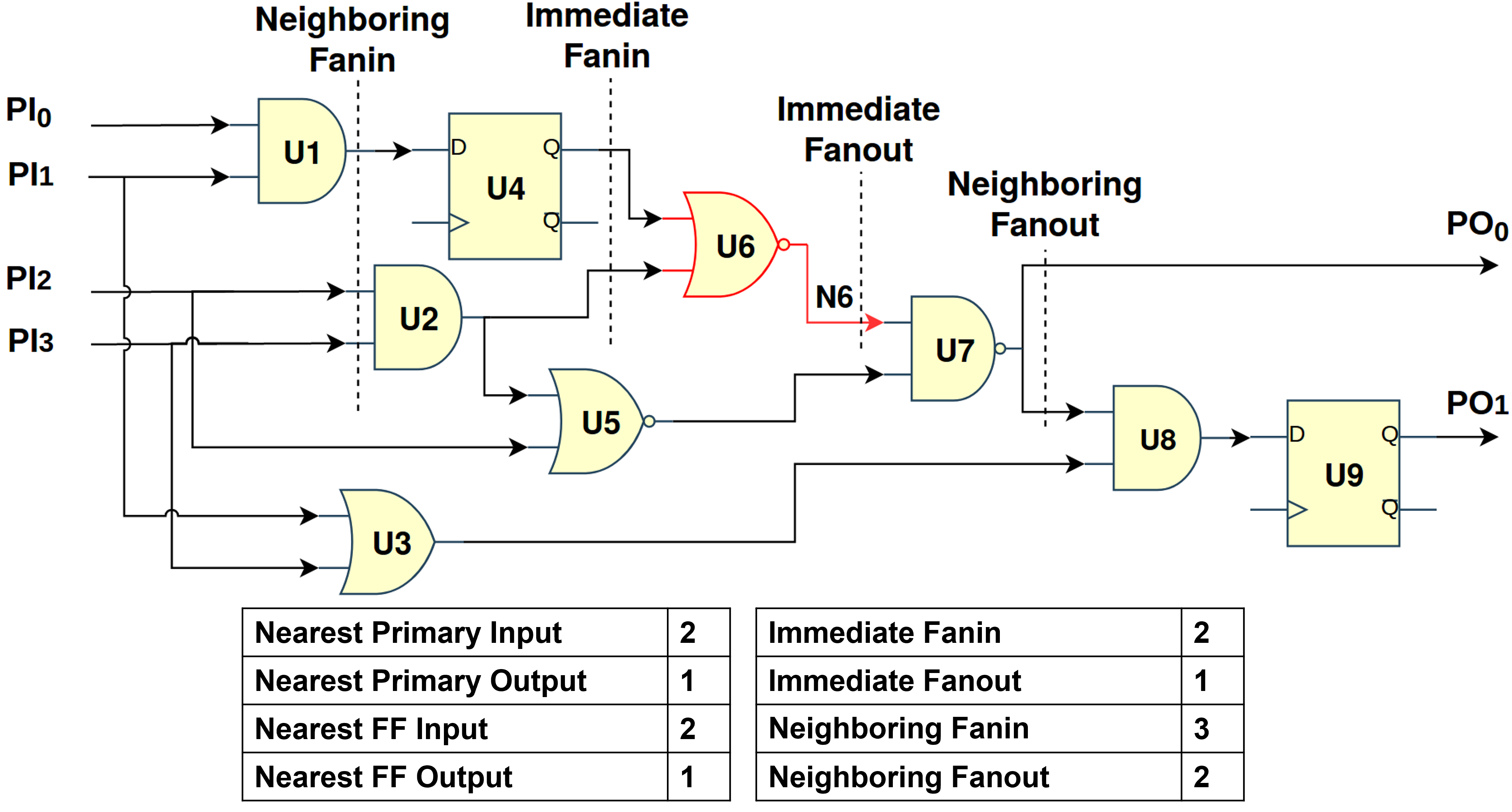}
	\caption{Illustration of structural features for the net $N6$ and its source gate $U6$ marked in red. } 
	\label{fig:struct}
	
\end{figure}

\begin{table*}[t!]
\centering
\caption{List of Net Features used in MIMIC}
\label{table:feature}
\begin{tabular}{|l|l|c|}  \hline
  \multicolumn{1}{|c|}{\textbf{Feature}} & \multicolumn{1}{c|}{\textbf{Description}} & \multicolumn{1}{c|}{\textbf{Type}}             \\ \hline
Signal Probability                         & Probability of the net to be logic 1         & Functional        \\ \hline
 Toggle Rate                           & Probability of the net to toggle its value    & Functional                 \\  \hline
Entropy                                     & Truth table of the cell driving the net      & Functional      \\  \hline 
 Control-to-0                             & Effort to control net to logic 0  & Functional                  \\  \hline
 Control-to-1                             & Effort to control net to logic 1  & Functional                  \\  \hline
Observability                               & Effort to propagate value of net to an observable point  & Functional                    \\\hline
Nearest Primary Input      & Minimum distance from a primary input to the net& Structural           \\ \hline
 Nearest Primary Output      & Minimum distance to a primary output from the net   & Structural       \\ \hline
  Immediate Fanin &  Number of immediate inputs to the driving gate & Structural   \\ \hline
  Immediate Fanout & Number of immediate outputs from the driving gate & Structural  \\ \hline
  Neighboring Fanin & Number of inputs to the driving gate beyond depth 1   & Structural          \\ \hline
  Neighboring Fanout & Number of outputs from the driving gate beyond depth 1   & Structural          \\ \hline
 Nearest Flip-Flop Input      & Minimum distance of the net to a flip-flop input & Structural\\ \hline
 Nearest Flip-Flop Output  &  Minimum distance of the net to a flip-flop output & Structural\\ \hline
 
\end{tabular}
\end{table*}
MIMIC first converts the flattened Verilog gate-level netlists to a hypergraph representation to obtain the necessary Trojan features for training the ML models.  Once in this form, we convert it to a directed acyclic graph assuming full-scan implementation, topologically sort the gates, and compute functional and structural features from every net in the design.  MIMIC automatically extracts these features from the sample Trojans for training as well as the target design for Trojan insertion. 
\cref{table:feature} lists all features we use in the current iteration which is inspired by \cite{hoque2018hardware}, with associated illustration in \cref{fig:struct}.  We refer to the collection of these features as \textit{net features}. These features can broadly be grouped into structural and functional net features.  Note that MIMIC has been designed to incorporate any number of provided features.  For self-containment, we will now describe the currently supported features in greater detail.  
\\\textbf{Functional Features} explain the functional behavior of the net or its input logic cone. Signal probability is a value from 0 to 1 calculated by the percentage of time a net is logic 1. The toggle rate is the percentage of time a net transitions from 0-to-1 or 1-to-0 over some time period. A Trojan is expected to be on either extreme of signal probability (hard-to-activate 1 or hard-to-activate 0) and exhibit a low toggle rate so as to avoid detection during logic testing.   Entropy is a measure of the balance of zeros and ones in the truth table of the source or driving gate. MIMIC calculates the entropy of the gate as $E=P_{1} \log_2 \dfrac{1} {P_{1}}  + P_{0} \log_2 \dfrac{1} {P_{0}}$, where $P_0, P_1$ are the probability of logic 0 and logic 1 calculated from the truth table of the source gate, respectively. We also incorporate Sandia Controllability and Observability (SCOAP) Values \cite{ref:scoap}. Controllability is determined by the effort in controlling a net to logic 0 or 1. Observability, on the other hand, is the effort required to propagate a signal to an observable point in the design, such as a primary output or scan flip-flop. Stealthy Trojans generally exhibit high SCOAP values which make them hard-to-control and/or hard-to-observe.
\\\textbf{Structural Features} describe information related to the connection of the nets and their topology. From \cref{table:feature}, we capture the shortest distance from primary inputs and outputs and flip flop data inputs and outputs.  Other structural features describe the fanin and fanout structure.  Immediate fanin and fanout pertain to the direct inputs and outputs of a source gate. Neighboring fan structures are the fanin and fanout values that are one depth forward and back from the source gate. An example calculation of structural features is shown in \cref{fig:struct}.  MIMIC calculates all structural features through breadth-first search and depth-first search graph traversal algorithms. Structural features provide insight into Trojan locations as well as common Trojan assets. For example, a Trojan that leaks information through function will most likely be close to a primary output or flip flop.

\subsubsection{Trigger and Payload Models}
The trigger and payload models are classifiers trained on feature vectors from the trigger and payload nets of sample hardware Trojans, respectively. Given a net from the target design (where we are to insert the Trojans), the trained models will be able to provide a score denoting the fitness of the said net for usage as a Trojan trigger or payload based on the sampled Trojan set. We defined this score as a function of the prediction probability or the decision boundary distance. We have evaluated different classification algorithms such as SVM, SVM One Class, and Random Forest (RF) and found that RF performs best for our experimental setup. Training a two-class classifier (such as RF) requires additional considerations in case of imbalanced training data (which may be the case in this task). Hence, through empirical analysis, we employ a balanced class weight scheme as described by sklearn~\cite{ref:sklearn}.

\subsubsection{Trojan Model}
 
We generate a set of virtual Trojans through a combination of the candidate trigger and payload nets with high fitness scores calculated based on the trigger and payload models. The virtual Trojans are the feature vectors calculated as if the Trojan were ``plugged into" the design. These virtual Trojans form a sub-space in the entire Trojan space that mimics the sample Trojan class. The function of the Trojan model is to rank these virtual Trojans based on their similarity with respect to the sample Trojans. The Trojan model is trained on the final trigger nets of the sample Trojans using 5 main functional features: probability, activity, control-to-1, control-to-0, and observability. We define this subset of features as \textit{Trojan features}.  We select the final trigger wire to define Trojan behavior as it is located at the critical junction between trigger and payload subcircuits, capturing the behavior of both.  To increase Trojan diversity but at the same time ensure coherence with the sample Trojan class, we realize the Trojan model as a generative machine learning model. Similarly, MIMIC can adapt to adjust the included features for the Trojan model.

\subsection{Operation Phase}
For Trojan insertion, MIMIC expects the previously described ML models for the trigger, payload, and Trojan, a flattened gate-level target netlist, and Trojan template netlist to be inserted. Algorithm \ref{algo:insert} details the steps followed to insert a Trojan netlist ($T$) into a provided target netlist ($V$) using three ML models for trigger ($ML_T$), payload ($ML_P$), and Trojan ($ML_{Troj}$) which describe the features of the trigger nets, payload nets, and Trojan behavior, respectively. The algorithm provides as output the Trojan inserted netlist ($V_{Troj}$) and a description of the inserted Trojans. In lines 1-4, MIMIC converts the input Verilog netlist and Trojan netlist to a hypergraph representation and performs a topological sort. After converting the input netlists to a graph, MIMIC extracts the features described above for every net in the supplied netlist and then scales their values (lines 6-11).  Once the features have been extracted and scaled, we can begin to identify suitable trigger and payload nets and then insert the Trojan.

\begin{algorithm}[h]
\begin{spacing}{1}
\caption{Trojan Insertion }
\label{algo:insert}
  \SetKwInOut{Input}{Input}
  \SetKwInOut{Output}{Output}
  \Input{Netlist ($V$),\\ ML Models ($ML_{T}, ML_{P}, ML_{Troj}$),\\ Trojan Netlist ($T$)}
  \Output{Trojan Inserted Netlist ($V_{Troj}$)}

  \textit{G} $\leftarrow$ \text{parse $V$ and construct hypergraph}\\
  \textit{G} $\leftarrow$ \text{topological sort $G$}\\
  \textit{GT} $\leftarrow$ \text{parse $T$ and construct hypergraph}\\
  \textit{GT} $\leftarrow$ \text{topological sort $GT$}\\
  \textit{r} $\leftarrow$ \text{number of trigger inputs of $GT$}\\
  $F_{N}$ $\leftarrow$  $\emptyset$  \\
  \ForEach{edge $\in$ $G$}{
    $f$ $\leftarrow$ \text{compute net feature set of $edge$} \\
   $F_N$ $\leftarrow$ $F_N \cup f$ \\
    }
   $F_N'$ $\leftarrow$ \text{scale $F_N$}\\
    $C_{trig}$  $\leftarrow$ \text{find candidate trigger nets with $ML_{T}$}\\
    $C_{trig}$ $\leftarrow$  \text{enumerate  $|C_{trig}| \choose r$}\\
   $C_{trig}$ $\leftarrow$ \text{validate candidate trigger sets in $C_{trig}$} \\
   \text{$C_{Troj}$} $\leftarrow$ $\emptyset$ \\
   \ForEach{trigger set x $\in C_{trig}$}{
     payload $\leftarrow$ \text{find candidate payload net with $ML_{P}$} \\
    $C_{Troj}$ $\leftarrow$ \text{($x$,payload)}$ \cup  $ $C_{Troj}$ \\
   }
    $C_{virt}$ $\leftarrow$ $\emptyset$ \\
   \ForEach{Trojan wire set x $\in$ $C_{Troj}$}
   {
    $GT_x$ $\leftarrow$ \text{plug in $x$ into $GT$}\\
    $F_T$ $\leftarrow$ $\emptyset$ \\
    \ForEach{edge $\in GT_x$ }
    {
         $f$ $\leftarrow$ \text{compute Trojan feature set of $edge$}\\
        $F_T$ $\leftarrow$ $F_T \cup f$ \\ 
    }
     $F_T'$ $\leftarrow$ \text{scale $F_N \cup F_T$}\\
     $C_{virt}$ $\leftarrow$ \text{ $F_T'$ $\cup$ $C_{virt}$} \\
   }
    $R_{Troj}$ $\leftarrow$ \text{sample feature set from $ML_{Troj}$} \\
   $C_{virt}$ $\leftarrow$ \text{sort $C_{virt}$ on distance from $R_{Troj}$}\\
    $V_{Troj}$ $\leftarrow$ \text{ $C_{virt}$ $ \cup$ $V_{Troj}$} \\

\textbf{return} $V_{Troj}$
\end{spacing}
\end{algorithm}

 \subsubsection{Selection of Trigger and Payload Nets}
With the scaled feature vector (line 11), we can begin to identify candidate nets using the trigger net ML model ($ML_T$) and payload net ML model. ($ML_P$). 
In line 12, we identify candidate trigger nets with $ML_T$.  The model is used to extract the probability estimates for the trigger class for each net using the scaled net feature vectors in $F_N'$.  In our case, the probability estimate is a two-element array of probabilities for normal net and trigger net classes, which sum to 1.  A net is classified as a trigger net if its trigger class probability estimate is greater than 0.5.  Therefore, nets with a satisfying probability estimate for trigger class are preferred. Only when there are an insufficient number of candidate nets will the algorithm select nets with a prediction probability less than 0.5. The nets are then sorted in descending order of probability estimates.  $r$-trigger sets are then generated in line 13.  To further increase the diversity of net selection, the algorithm selects a net with a prediction probability different than the previously selected net until $r$ unique nets have been identified.   By selecting nets with features similar to the reference features as the trigger, MIMIC ensures the generated Trojans have characteristics that mimics sampled Trojans used for training.

Once the trigger net sets have been created, we need to check whether these trigger nets can achieve the combination of values required to activate the Trojan (line 14). Failure to perform this check will lead to the formation of ``dead" Trojans (Trojans present in the netlist but never trigger). 
In our case, we have support for Cadence JasperGold, Synosys TetraMAX, and Onespin to perform this check. Most formal or automatic test pattern generation (ATPG) computer-aided design (CAD) tools can be used to achieve similar results.  If every trigger net can simultaneously achieve the required value, the selection of nets is deemed valid, and MIMIC can proceed to payload selection.  
Similar to the trigger selection process, the payloads are selected with the use of the payload model. Payloads are rank-ordered according to their prediction probability for payload net class. In addition to the previous qualifiers for trigger selection, payloads are selected such that they do not form a combinational loop (line 17). This property is enforced by selecting a payload that has a higher topological ordering than the maximum topological order from the set of trigger nets.  With a set of validated triggers and payload nets (line 18), MIMIC moves onto Trojan insertion.

\subsubsection{Trojan Insertion}
With a set of valid trigger and payload nets, MIMIC begins the Trojan insertion stage.
From this set, MIMIC creates virtual Trojans -- the feature vector associated with Trojan features -- by calculating the Trojan features as if the Trojan were inserted into the design (lines 21-30).  
The virtual Trojans are then sorted in ascending order based on the Euclidean distance from the ideal feature vector, which is sampled from the Trojan Generative Model, $ML_{Troj}$ (lines 31-32).  MIMIC has the capability of generating several virtual Trojans, evaluating their fitness, and ranking their closeness to the provided generative model.  MIMIC will select from the fittest virtual Trojans, and insert them into the design graph (line 33). A report is also generated that details information on where the Trojan is inserted and the corresponding Trojan features.

\section{Results}
\label{results}

\begin{figure*}[t!]
 	\centering
    \includegraphics[width=\textwidth]{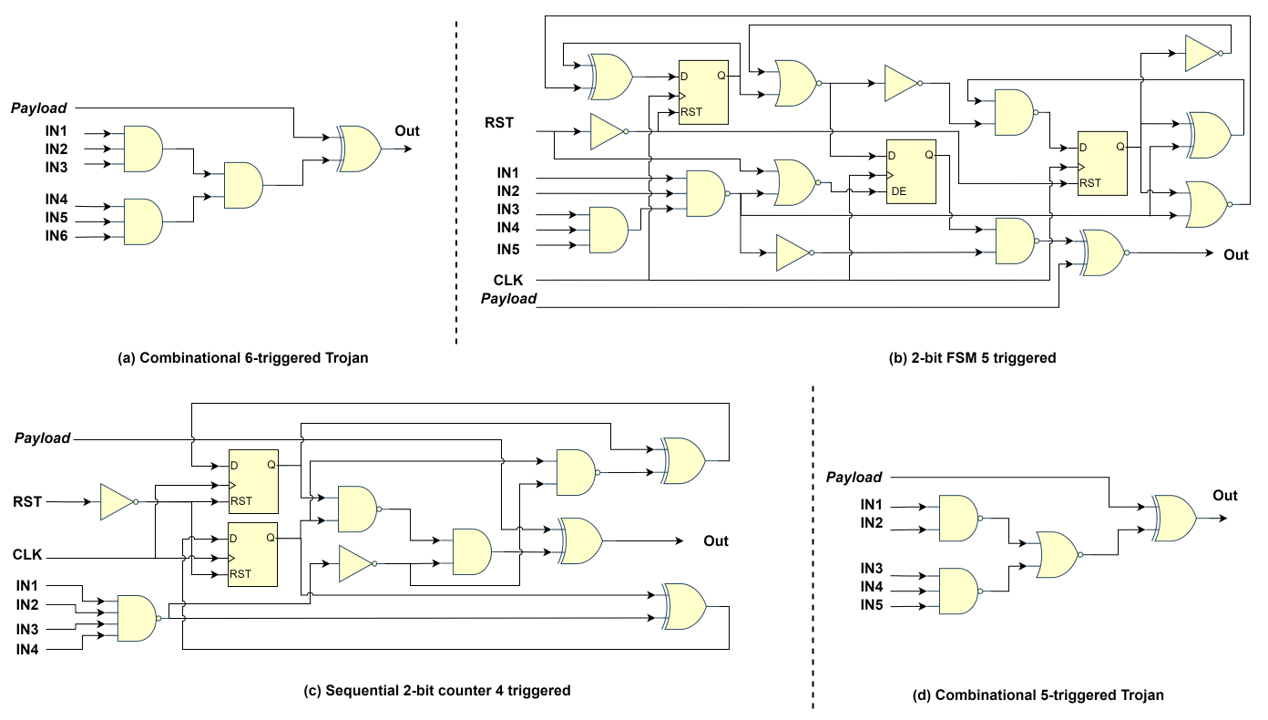}
 	\caption{Denial-of-Service Trojan Templates used during MIMIC insertion. (a) 6 trigger combinational DoS, (b) 5 trigger 2-bit FSM  DoS, (c) 4 trigger 2-bit counter DoS, (d) 5 trigger combinational DoS. }
 	 \label{fig:dc_trojan}
 
 \end{figure*}
A powerful feature of the MIMIC tool is to automatically generate similar Trojans to those that were used for training.  In this section, we perform several experiments across four ISCAS 89 benchmarks (s5378, s9234, s38417, s38584) and four Trojan templates shown in \cref{fig:dc_trojan} to highlight the effectiveness of MIMIC in learning and reproducing similar Trojans from a dataset.

\subsection{Experimental Setup}
\begin{figure*}[t!]
 	\centering
 	\includegraphics[width=\textwidth]{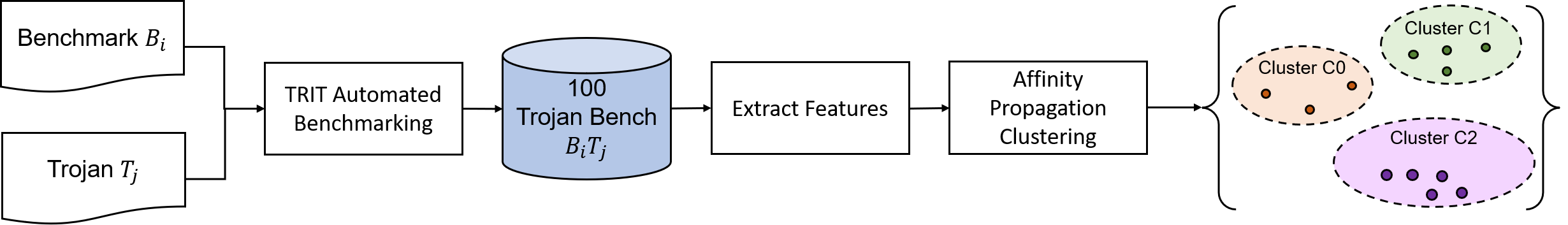}
 	\caption{Experimental setup flow for producing Trojans to train and evaluate MIMIC.}
 	\label{fig:experimental_setup}
 
 \end{figure*}
For each benchmark ($B_i$) - Trojan template ($T_j$) pair, we create 100 Trojan inserted benchmarks ($B_iT_j$) using a flow similar to \cite{ref:trit}. The following parameters are used given the number of trigger inputs: 4-trigger, $\theta$=0.001; 5-trigger, $\theta$=0.01; 6-trigger, $\theta$=0.01; where $\theta$ is the signal probability threshold for selecting trigger nets. Both net and Trojan features are extracted for all nets in the Trojan inserted benchmarks. We then use Affinity Propagation to cluster the Trojans within each set of 100 Trojan inserted benchmarks ($B_iT_j$) to form subsets of similar Trojans. Affinity Propagation will automatically generate a number of clusters based on the provided dataset and selected hyperparameters (damping=0.8, the remaining parameters are default~\cite{ref:sklearn}). For each resulting Trojan subset, we extract the trigger nets and payloads nets used for the Trojans and train a Random Forest classifier for trigger and payload models with net feature vector. This flow is illustrated in \cref{fig:experimental_setup} for clarification. Finally, each Trojan subset is used to train a Bayesian Gaussian Mixture model with the Trojan feature vector. In the end, for every Trojan cluster within a benchmark - Trojan template pair ($B_iT_j$), we have a trigger ML model, payload ML model, and Trojan ML Model.  These models are used for the subsequent experiments performed.
 \subsection{Same Template, Same Benchmark (STSB)}

\begin{table*}[h!]
\centering
\caption{Evaluation of MIMIC under Same Template, Same Benchmark Scenario using Structural \& Functional Features}
\label{results:STSB}
\begin{tabularx}{\textwidth}{c|c|Y|Y|Y|Y|Y|Y|Y|Y}
\cline{1-10} 
\multirow{2}{*}{\textbf{Benchmark} } &\multicolumn{1}{c|}{\multirow{2}{*}{\begin{tabular}[c]{@{}c@{}}\textbf{Num}\\ \textbf{Clusters}\end{tabular}}} 
& \multicolumn{2}{c|}{\begin{tabular}[c]{@{}c@{}}\textbf{No ML}\\ \textbf{Acc.(\%)}\end{tabular}}
& \multicolumn{2}{c|}{\begin{tabular}[c]{@{}c@{}}\textbf{Troj. ML (A)}\\ \textbf{ Acc.(\%)}\end{tabular}} 
& \multicolumn{2}{c|}{\begin{tabular}[c]{@{}c@{}}\textbf{Trig.\&Pay. ML (B)}\\ \textbf{ Acc.(\%)}\end{tabular}} 
& \multicolumn{2}{c}{\begin{tabular}[c]{@{}c@{}}\textbf{Both (A) and (B)}\\\textbf{Acc.(\%)}\end{tabular}} \\ \cline{3-10} 
                                 &  & \textbf{Top-1}    & \textbf{Top-5}   & \textbf{Top-1 }   & \textbf{Top-5} & \textbf{Top-1}   & \textbf{Top-5} & \textbf{Top-1}  & \textbf{Top-5} \\ \cline{1-10}
s5378-c1   & 14                 & 2.86        & 4.29       & 15.71        & 15.71       & 4.29       & 17.14 & 62.86 & 64.29      \\ 
s5378-c2   & 12                & 0.00        & 5.00       & 15.00        & 15.00       & 1.67       & 18.33  & 60.00 & 61.67    \\ 
s5378-s1   & 12                & 0.00        & 1.67       & 15.00        & 15.00       & 6.67       & 23.33   & 81.67 & 85.00    \\ 
s5378-s2   & 8                & 2.50        & 5.00       & 25.00        & 25.00       & 5.00       & 20.00   & 72.50 & 75.00    \\ \cline{1-10}
s9234-c1  & 10                 & 4.00        & 8.00       & 40.00        & 40.00       & 2.00       & 16.00   & 56.00 & 60.00    \\ 
s9234-c2  & 6                 & 3.33        & 10.00       & 16.67        & 16.67       & 10.00       & 16.67  & 73.33 & 76.67     \\ 
s9234-s1  & 11                 & 1.82        & 3.64       & 18.18        & 20.00       & 3.64       & 18.18  & 74.55 & 76.36     \\ 
s9234-s2  & 6                 & 3.33        & 13.33       & 23.33        & 30.00       & 3.33       & 26.67   & 80.00 & 83.33    \\ \cline{1-10}
s38417-c1  & 6                & 6.67        & 10.00       & 46.67        & 46.67       & 3.33       & 26.67  & 96.67 & 100.00     \\ 
s38417-c2  & 6                & 0.00        & 6.67       & 23.33        & 30.00       & 0.00       & 36.67 & 93.33 & 100.00     \\ 
s38417-s1  & 9                & 2.22        & 4.44      & 28.89        & 28.89       & 2.22       & 13.33   & 64.44 & 64.44    \\ 
s38417-s2  & 9                & 2.22        & 15.56       & 44.44        & 46.67       & 8.89       & 26.67    & 86.67 & 86.67   \\ \cline{1-10}
s38584-c1  & 8                & 0.00        & 0.00       & 15.00        & 15.00       & 7.50       & 32.50  & 80.00 & 87.50     \\ 
s38584-c2  & 8                & 0.00        & 2.50       & 22.5        & 25.00       & 5.00       & 22.50   & 75.00 & 85.00    \\ 
s38584-s1  & 8                & 0.00        & 2.50       & 17.50        & 17.50       & 5.00       & 27.50  & 97.50 & 100.00     \\ 
s38584-s2  & 9                & 0.00        & 0.00       & 17.78        & 17.78       & 2.22       & 6.67  & 64.44 & 66.67     \\ \cline{1-10}
Average & -- & 1.81 & 5.79 & 24.07 & 25.31 & 4.42 & 21.80 & 76.18 & 79.54\\ \cline{1-10}
\end{tabularx}
\\ Trig=Trigger; Pay=Payload; Troj=Trojan; Acc.=Accuracy; (A) uses only Trojan ML; (B) uses only Trigger \& Payload ML;
\end{table*}

 At the most basic level, we want to evaluate the MIMIC framework's success in producing different but similar quality Trojans to those used for training. In effect, MIMIC is providing alternative suitable locations that will provide similar Trojan qualities for a given Trojan template in a benchmark. For these experiments, the template and benchmark for insertion are the same benchmark and Trojan template used for training the ML models (hence the name Same Template, Same Benchmark (STSB)).  We separately mimic each Trojan cluster and try to produce Trojans that fall in the correct cluster.  We maintain a sample of 20 virtual Trojans per request of 1 Trojan.  We then run MIMIC 5 times for each benchmark - Trojan template pair to better understand the performance.
In \cref{results:STSB}, the first column represents the benchmark and Trojan template used, separated by ``-''. The second column reports the number of Trojan clusters generated from Affinity Propagation.  To highlight the benefits of the ML components, we compare the accuracy of MIMIC for several scenarios with and without the Trigger and Payload Models and Trojan Models (columns 3-10). The column names denote the inclusion and exclusion of a model. For example, No ML (column 3) includes no models, Trojan ML (A) (column 4) includes only the Trojan Model, etc. The accuracy is evaluated by identifying the corresponding cluster of the output Trojan. The output Trojan is considered correct if it is successfully placed in the same cluster used for training.   Because MIMIC maintains a sorted list of several ``virtual'' Trojans that have not been inserted into the target benchmark, we report accuracy in the form of Top-$N$.  Top-$N$ accuracy refers to finding a correctly clustered Trojan within the top $N$ virtual Trojans sorted on the distance from the reference Trojan features from the Trojan model.  If the Trojan model is excluded, the first $N$ Trojans are selected without sorting from the Trojan Model. If the trigger and payload models are excluded, the nets are randomly selected but still form a valid trigger and payload set.  From \cref{results:STSB}, we observe an increase in the accuracy of the framework with the inclusion of each model individually. There is an increase of approximately 16\% and 20\% Top-5 accuracy with the inclusion of the trigger and payload models (column 8), and Trojan models (column 6), respectively. With the trigger, payload, and Trojan models included, we observe around 80\% Top-5 accuracy or an increase of around 74\%  over the accuracy without any models.  Ultimately, from \cref{results:STSB}, MIMIC is shown to properly identify other suitable locations for Trojan templates that exhibit the same behavior as the Trojans used for training.

\subsection{Extended Use Cases}
We further study MIMIC's flexibility with a more realistic training-testing scenario by testing on the same benchmark but inserting a different template. This experiment is evaluated using all ML models (trigger, payload, and Trojan Models) trained using only functional features. 

\subsubsection{Different Template, Same Benchmark (DTSB)}
\begin{figure*}[h]
  	\centering
  	\includegraphics[width=\textwidth]{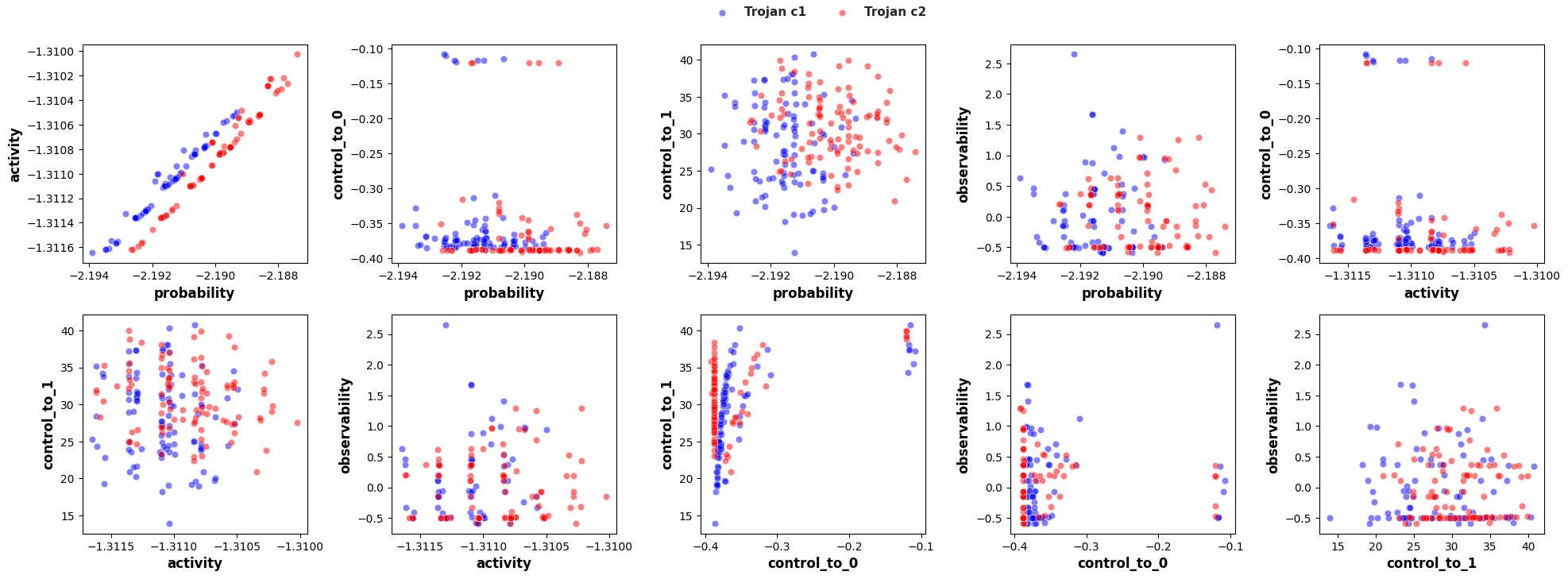}
  	\caption{Comparing pair-wise functional features of combinational DoS Trojan template c1 and c2 inserted in s38417.}
  	\label{fig:DTSB1}
  \end{figure*}

\begin{figure*}[h]
  	\centering
  	\includegraphics[width=\textwidth]{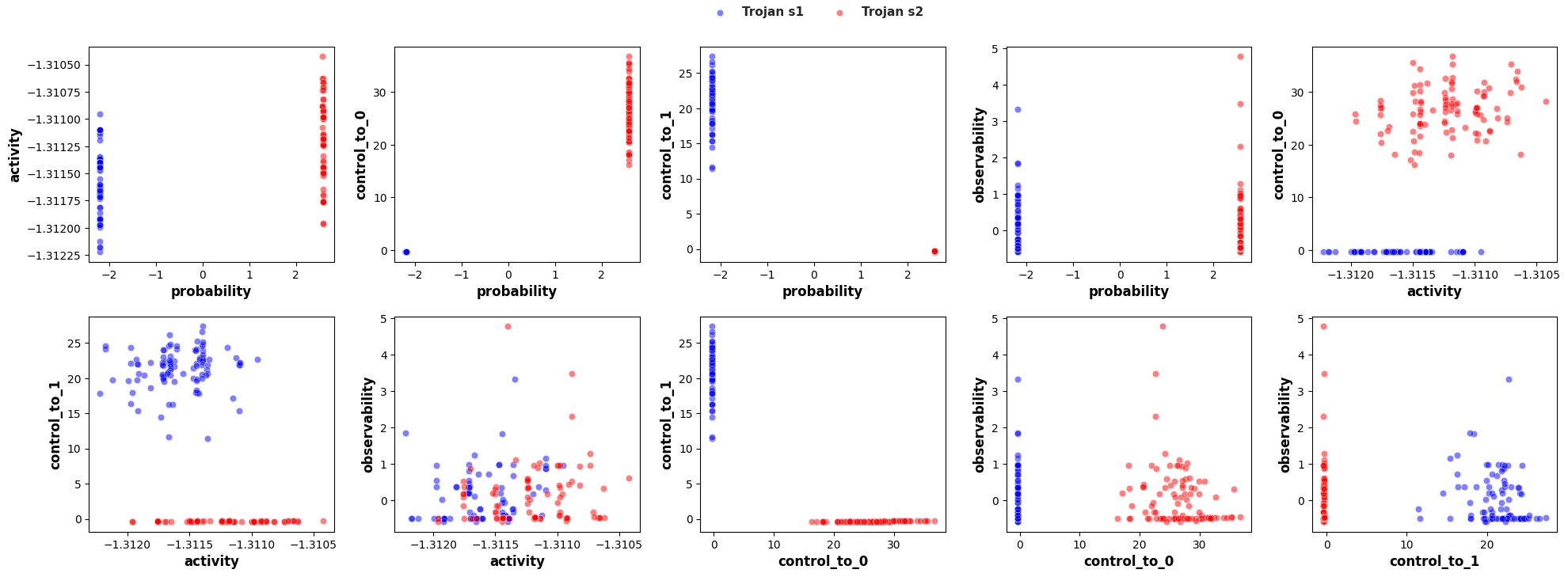}
  	\caption{Comparing pair-wise functional features of sequential DoS Trojan template s1 and s2 inserted in s38417.}
  	\label{fig:DTSB2}
  \end{figure*}

For DTSB, we want to evaluate if MIMIC can find suitable locations for a different Trojan template in the same benchmark to achieve the same Trojan behavior of the original Trojan template used for training.  Hence, we use the models trained for a specific benchmark-Trojan template pair to insert a different Trojan template into the same benchmark used during training.  In \cref{results:DTSB}, the first column represents the benchmark and Trojan used for insertion.  MIMIC can achieve good accuracy between Trojan templates c1 and c2.  We see an average Top-5 accuracy of around 87\% for c1-train, c2-test and around 77\% average accuracy for c2-train, c1-test. \cref{fig:DTSB1} plots the pair-wise functional features of combinational Trojan templates c1 and c2. From these feature distributions, we can see there is significant overlap, which indicates similar Trojan behavior. Conversely, \cref{fig:DTSB2} illustrates the pair-wise functional features of the sequential templates s1 and s2.  We can clearly observe templates s1 and s2 do not share much feature space in common. This disparity is reflected in the poor performance between templates s1 and s2 shown in \cref{results:DTSB}.  MIMIC is unable to find suitable locations for Trojan template s1 to match the behavior of the Trojan template s2, and vice versa. Note, we do not test across combinational and sequential template boundaries as the presence of flip flops in the sequential Trojan templates heavily affects the relevant feature space. 
\begin{table}[h!]
\centering
\caption{Evaluation of MIMIC under Different Template, Same Benchmark Scenario using Functional Features}
\label{results:DTSB}
\begin{tabularx}{0.5\textwidth}{c|c|c|c|c}
\hline
\multirow{2}{*}{\textbf{Benchmark}}  &\multicolumn{1}{c|}{\multirow{2}{*}{\begin{tabular}[c]{@{}c@{}}\textbf{Training} \\ \textbf{Benchmark}\end{tabular}}} &\multicolumn{1}{c|}{\multirow{2}{*}{\begin{tabular}[c]{@{}c@{}}\textbf{Training} \\ \textbf{Template}\end{tabular}}}& \multicolumn{2}{c}{\textbf{Accuracy(\%)}} \\ \cline{4-5} 
                                    & & & \textbf{Top-1}    & \textbf{Top-5}             \\ \hline
s9234-c1     & s9234 & c2               & 58.00        & 60.00                  \\ 
s9234-c2     & s9234 & c1              & 73.33        & 80.00              \\ 
s9234-s1     & s9234 & s2              & 9.09        & 9.09                   \\ 
s9234-s2     & s9234 & s1              & 30.00        & 30.00                   \\ 
\hline\hline
s38417-c1     & s38417 & c2               & 90.00        & 93.33             \\ 
s38417-c2     & s38417 & c1              & 90.00        & 93.33                 \\ 
s38417-s1     & s38417 & s2              & 11.11        & 11.11                    \\ 
s38417-s2     & s38417 & s1              & 11.11        & 11.11                   \\ \hline
\end{tabularx}
\end{table}

\section{Conclusion}
\label{conclusion}
We have presented MIMIC, a machine learning-based CAD framework for automated hardware Trojan insertion to effectively explore the attack surface for a design and generate a large pool of high-quality synthetic benchmarks. The need for unbiased automated Trojan benchmarking has become apparent as more countermeasures for hardware Trojan attacks are being developed. Contrary to existing Trojan insertion techniques, MIMIC aims to learn several key features of hard-to-detect Trojans to create stealthy Trojans of similar nature. It offers the flexibility of several user configurations, which can control the quality and types of Trojans inserted into a gate-level design. We have developed a complete tool flow for MIMIC and demonstrated its ability to mimic Trojan properties through several experiments using the same benchmark and Trojan template and also differing Trojan templates. We have described possible use cases of MIMIC in a trustworthy hardware design life cycle. Based on these use cases, we can expect that Trojan insertion frameworks, such as MIMIC, will play an increasingly important role in quantifiable assurance for electronic hardware in the emergent business model. Future work will investigate further the evaluation of MIMIC in specific use cases and for emerging Trojan attacks, as well as feedback-based insertion and automatic template selection based on the input models.

\bibliographystyle{IEEEtran}
\bibliography{IEEEexample}

\end{document}